\documentclass[preprint2]{aastex}
 \usepackage[fleqn]{amsmath}
 \usepackage{graphicx}
\usepackage{dcolumn}
\usepackage{bm}
\usepackage{multirow}

\shorttitle{Construction of explicit symplectic integrators}
\shortauthors{Wang et al.}

\begin{document}


\title{Construction of explicit symplectic integrators in general relativity. I. Schwarzschild black holes}


\author{Ying Wang$^{1,2}$, Wei Sun$^{1}$, Fuyao Liu$^{1}$, Xin Wu$^{1,2,3,\dag}$}
\affil{1. School of Mathematics, Physics and Statistics, Shanghai
University of Engineering Science, Shanghai 201620, China
\\ 2. Center of Application and Research of Computational Physics,
Shanghai University of Engineering Science, Shanghai 201620, China
\\  3. Guangxi Key Laboratory for Relativistic Astrophysics, Guangxi
University, Nanning 530004, China} \email{Emails:
wangying424524@163.com (Y. W.), sunweiay@163.com (W. S.),
liufuyao2017@163.com (F. L.); $\dag$ Corresponding Author:
wuxin$\_$1134@sina.com (X. W.)}


\begin{abstract}

Symplectic integrators that preserve the geometric structure of
Hamiltonian flows and do not exhibit secular growth in energy
errors are suitable for the long-term integration of N-body
Hamiltonian systems in the solar system. However, the construction
of explicit symplectic integrators is frequently difficult in
general relativity  because all variables are inseparable.
Moreover, even if two analytically integrable splitting parts
exist in a relativistic Hamiltonian, all analytical solutions are
not explicit functions of proper time. Naturally, implicit
symplectic integrators, such as the midpoint rule, are applicable
to this case. In general, these integrators are numerically more
expensive to solve than  same-order explicit symplectic
algorithms. To address this issue, we split the Hamiltonian of
Schwarzschild space-time geometry into four integrable parts with
analytical solutions as explicit functions of proper time. In this
manner, second- and fourth-order explicit symplectic integrators
can be easily available. The new algorithms are also useful for
modeling the chaotic motion of charged particles around a black
hole with an external magnetic field. They demonstrate excellent
long-term performance in maintaining bounded Hamiltonian errors
and saving computational cost when appropriate proper time steps
are adopted.

\end{abstract}


\emph{Unified Astronomy Thesaurus concepts}: Black hole physics
(159); Computational methods (1965); Computational astronomy
(293); Chaos (222)




\section{Introduction}
\label{sec:intro}

Black holes and gravitational waves were predicted in Einstein's
theory of general relativity (Einstein 1915; Einstein $\&$
Sitzungsber 1916). The Schwarzschild solution was obtained from
the field equations of a nonrotating black hole (Schwarzschild
1916). The Kerr solution was given to a rotating black hole (Kerr
1963). The recent detection of gravitational waves (GW150914) from
a binary black hole merger (Abbott et al. 2016) and the images of
a supermassive black hole candidate at the center of the giant
elliptical galaxy M87 (EHT Collaboration et al. 2019) provide
powerful evidence for confirming the two predictions.

Although the relativistic equations of motion for test particles
in the Schwarzschild and Kerr metrics are highly nonlinear, they
are separable in variables and solved analytically in the
Hamiltonian-Jacobi equation. Thus, they are integrable and the
motions of particles near the two black holes are strictly
regular. This integrability is attributed to the existence of four
independent constants of motion, namely, energy, angular momentum,
four-velocity relation of particles, and Carter constant (Carter
1968). However, no additional information regarding the solutions
but only the integrability of space-times is known because the
solutions are expressed in terms of quadratures rather than
elementary functions. Good numerical methods for computing these
geodesics are highly desirable. In particular, when magnetic
fields are included in curved space-times, the separation of
variables in the Hamiltonian-Jacobi equation, associated to the
equations of charged particle motion, is generally highly
improbable. This condition may lead to the non-integrability of
systems and the chaotic behavior of motion (Takahashi $\&$ Koyama
2009; Kop\'{a}\v{c}ek et al. 2010; Kop\'{a}\v{c}ek $\&$ Karas
2014; Kolo\v{s} et al. 2015; Stuchl\'{i}k $\&$ Kolo\v{s} 2016;
Tursunov et al. 2016; Azreg-A\"{i}nou 2016; Li $\&$ Wu 2019).
Numerical methods play an important role in analyzing the
properties of these non-integrable problems.

Supposedly, good numerical methods are integrators that provide
reliable results, particularly in the case of long-term
integrations. In addition, the preservation of structural
properties, such as symplectic structures, integrals of motion,
phase-space volume and symmetries, should be desired. Such
structure-preserving algorithms belong to a class of geometric
integrators (Hairer et al. 1999). Among the properties,  the most
important ones are the preservation of energy and symplecticity.

In many cases, checking energy accuracy is a basic reference for
testing the performance of numerical integration algorithms
although energy conservation does not necessarily yield
high-precision numerical solutions. To demonstrate this scenario,
we present a two-body problem as an example. Energy errors from
the truncation or discretization errors of Runge-Kutta type
algorithms in the two-body problem typically increase linearly
with integration time (Rein $\&$ Spiegel 2015). The growth speeds
of in-track errors (Huang $\&$ Innanen 1983), which correspond to
errors along the tangent to a trajectory in phase space, directly
depend on the relative error in Keplerian energy (Avdyushev 2003).
Accordingly, the Keplerian orbit is Lyapunov's instability that
leads to an increase in various errors. However, the stabilization
or conservation of energy along the orbit is more efficient in
eliminating Lyapunov's instability and the fast drifting of
in-track errors than that of other integrals. The energy
stabilization method of Baumgarte (1972, 1973) includes known
integrals (such as an energy integral) in the equations of motion.
The stabilization in the perturbed two-body, restricted three-body
problems of satellites, asteroids, stars and planets has been
demonstrated to improve the accuracy of numerical integrations by
several orders of magnitude (Avdyushev 2003). In contrast with
Baumgarte's method, the manifold correction or projection method
of Nacozy (1971) applies a least-squares procedure to add a linear
correction vector to a numerical solution. This vector is computed
from the gradient vectors of the integrals involving the total
energy. The application of Nacozy's method is generalized to
quasi-Keplerian motions of perturbed two-body or $N$-body problems
with the aid of the integral invariant relation of slowly varying
individual Kepler energies (Wu et al. 2007; Ma et al. 2008a). Some
projection methods (Fukushima 2003a, 2003b, 2003c, 2004; Ma et al.
2008b; Wang et al. 2016, 2018; Deng et al. 2020) for rigorously
satisfying integrals, including Kepler energy in a two-body
problem, have been proposed and extended to perturbed two-body
problems, $N$-body systems, nonconservative elliptic restricted
three-body problems and dissipative circular restricted three-body
problems. In addition to explicit projection methods that exactly
preserve the energy integral, exact energy-preserving implicit
integration methods that discretize Hamiltonian gradients in terms
of the average Hamiltonian difference terms have been specifically
designed for conservative Hamiltonian systems (Feng $\&$ Qin 2009;
Bacchini et al. 2018a, 2018b; Hu et al. 2019).

Although energy-preserving integrators and some projection methods
exactly  conserve energy, they are non-symplectic. Symplectic
algorithms (Wisdom 1982; Ruth 1983; Feng 1986; Suzuki 1991;
McLachlan $\&$ Atela 1992; Chin 1997; Omelyan et al. 2002a, 2002b,
2003) do not exactly conserve the energy of a Hamiltonian system,
but they cause energy errors to oscillate and become bounded as
evolution time increases. In this manner, these algorithms are
also considered to conserve energy efficiently over long-term
integrations. Moreover, they preserve the symplectic structure of
Hamiltonian flows. Given the two advantages, symplectic
integrators are widely used in long-term studies on solar system
dynamics. The most popular algorithms in solar system dynamics are
the second-order symplectic integrator of Wisdom $\&$ Holman
(1991) and its extensions (Wisdom et al. 1996; Chambers $\&$
Murison 2000; Laskar $\&$ Robutel 2001; Hernandez $\&$ Dehnen
2017). Notably, the explicit symplectic algorithms in a series of
references (Suzuki 1991; Chin 1997; Omelyan et al. 2002a, 2002b,
2003) require the integrated Hamiltonian to be split into two
parts with analytical solutions as explicit functions of time.
However, the two splitting parts from the Hamiltonian in Wisdom
$\&$ Holman (1991), Wisdom et al. (1996), Chambers $\&$ Murison
(2000) and Laskar $\&$ Robutel (2001) should be the primary and
secondary parts. For the secondary part, the analytical solutions
can be given in explicit functions of time. The primary part also
has explicit analytical solutions, but eccentric anomaly is
calculated using an iteration method, such as the Newton-Raphson
method.

However, a relativistic gravitational Hamiltonian system, such as
the Schwarzschild space-time, is inseparable or has no two
separable parts with analytical solutions being explicit functions
of proper time. This condition leads to the difficulty in applying
explicit symplectic integrators.  By extending the phase space of
such an inseparable Hamiltonian system,  Pihajoki (2015) obtained
a new Hamiltonian consisting of two sub-Hamiltonians equal to the
original Hamiltonian, where one sub-Hamiltonian is a function of
the original coordinates and new momenta, and the other is a
function of the original momenta and new coordinates. The two
sub-Hamiltonians are separable in variables; therefore, standard
explicit symplectic leapfrog splitting methods are applicable to
the new Hamiltonian. Mixing maps of feedback between the two
sub-Hamiltonian solutions and a map for projecting a vector in the
extended phase space back to the original number of dimensions are
necessary and have a suitable choice. Liu et al. (2016) confirmed
that sequent permutations of coordinates and momenta achieve good
results in preserving the original Hamiltonian without an increase
in secular errors compared with the permutations of momenta
suggested by Pihajoki (2015). Luo et al. (2017) found that
midpoint permutations exhibit the best results. However, mixing
maps generally destroy symplecticity in extended phase space. In
addition, extended phase space leapfrogs are not symplectic for
the use of any projection map. Despite the absence of
symplecticity, mixing and projection maps are used only as output
and exert no influence on the state in extended phase space.
Consequently, leapfrogs, such as partitioned multistep methods,
can exhibit good long-term behavior in stabilizing the original
Hamiltonian (Liu et al. 2017; Luo $\&$ Wu 2017;  Wu $\&$ Wu 2018).
Thus, extended phase-space leapfrog methods, including extended
phase-space logarithmic Hamiltonian methods (Li $\&$ Wu 2017), are
called explicit symplectic-like integrators. In addition to the
two copies of the original system with mixed-up positions and
momenta, a third sub-Hamiltonian, as an artificial restraint to
the divergence between the original  and extended variables, was
introduced by Tao (2016). Neither mixing nor projection maps are
used in Tao's method, and thus, explicit leapfrog methods are
still symplectic in the extended phase space. Two problems exist.
(\emph{i}) A binding constant for controlling divergence has an
optimal choice. This choice cannot be given theoretically but
requires considerable values to test which one minimizes the
original Hamiltonian error. (\emph{ii}) Whether the original
variables in the newly extended Hamiltonian coincide with those in
the original Hamiltonian is unclear.

To date, no standard explicit symplectic leapfrogs but only
implicit symplectic methods have been established in a
relativistic Hamiltonian problem because of the difficulty in
separating variables. The second-order implicit midpoint method
(Feng 1986) is the most common choice among implicit symplectic
methods. It can function as a variational symplectic integrator
for constrained Hamiltonian systems (Brown 2006). To save
computational cost, explicit and implicit combined symplectic
algorithms have been provided in some references (Liao 1997; Preto
$\&$ Saha 2009; Lubich et al. 2010; Zhong et al. 2010; Mei et al.
2013a, 2013b). Notably, the symplectic integration scheme for the
post-Newtonian motion of a spinning black hole binary (Lubich et
al. 2010) is noncanonical because of the use of noncanonical spin
variables. However, this scheme can become canonical when
canonically conjugated cylindrical-like spin coordinates (Wu $\&$
Xie 2010) are used. The symplectic implicit Gauss-Legendre
Runge-Kutta method has been applied to determine the regular and
chaotic behavior of charged particles around a Kerr black hole
immersed in a weak, asymptotically uniform magnetic field
(Kop\'{a}\v{c}ek et al. 2010). Implicit symmetric schemes with
adaptive step size control that effectively conserve the integrals
of motion are appropriate for studying geodesic orbits in curved
space-time backgrounds (Seyrich $\&$ Lukes-Gerakopoulos 2012).
Slimplectic integrators for general nonconservative systems (Tsang
et al. 2015) can share many benefits of traditional symplectic
integrators.

In general, implicit symplectic methods are numerically more
expensive to solve than same-order explicit symplectic
integrators. The latter algorithms should be used if possible.
Accordingly, we intend to address the difficulty in constructing
explicit symplectic integrators for Schwarzschild type space-times
similar to the standard explicit symplectic leapfrogs for
Hamiltonian problems in solar system dynamics. If the Hamiltonians
of Schwarzschild type space-times are separated into two parts
that resemble the splitting form of Hamiltonian systems in the
construction of standard symplectic leapfrogs, then no explicit
symplectic algorithms are available. The conditions for
constructing explicit symplectic schemes may require Hamiltonians
to be split into more parts with analytical solutions as explicit
functions of proper time.

The remainder of this paper is organized as follows. In Section 2,
we briefly introduce the standard explicit symplectic leapfrog and
its extensions for a separable Hamiltonian system. The Hamiltonian
of charged particles moving around a Schwarzschild black hole with
an external magnetic field is described in Section 3. Explicit
symplectic schemes are designed for curved Schwarzschild
space-times in Section 4. The performance of explicit symplectic
integrators is tested numerically in Section 5. Section 6
concludes the major results. A discrete difference scheme of the
new second-order explicit symplectic integrator is presented in
Appendix A. Explicit and implicit combined symplectic  methods and
extended phase-space explicit symplectic-like methods are provided
in Appendix B.

\section{Standard explicit symplectic integrators for a separable Hamiltonian}

Set $\mathbf{q}$ as an $N$-dimensional coordinate vector. Its
corresponding generalized momentum is $\mathbf{p}$. Let
$\mathbf{Z}=(\mathbf{p},\mathbf{q})$ be a $2N$-dimensional
phase-space variable. Consider the following Hamiltonian
\begin{equation}
H(\mathbf{p},\mathbf{q})=H_1(\mathbf{p},\mathbf{q})+H_2(\mathbf{p},\mathbf{q}),
\end{equation}
where the two separable parts $H_1$ and $H_2$ are supposed to be
independently integrable. A typical splitting form of $H$ takes
$H_1$ as kinetic energy $T(\mathbf{p})$ and $H_2$ as potential
$V(\mathbf{q})$.

Two differential operators are defined as follows:
\begin{eqnarray}
\mathcal{A} &=&\sum^{N}_{i=1}(\frac{\partial H_1}{\partial
\mathbf{p}_i}\frac{\partial }{\partial
\mathbf{q}_i}-\frac{\partial H_1}{\partial
\mathbf{q}_i}\frac{\partial }{\partial \mathbf{p}_i}), \nonumber
\\ \mathcal{B} &=&\sum^{N}_{i=1}(\frac{\partial H_2}{\partial
\mathbf{p}_i}\frac{\partial }{\partial
\mathbf{q}_i}-\frac{\partial H_2}{\partial
\mathbf{q}_i}\frac{\partial }{\partial \mathbf{p}_i}). \nonumber
\end{eqnarray}
System (1) has the following formal solution
\begin{equation}
\mathbf{Z}(h)=\mathcal{C}(h)\mathbf{Z}(0),
\end{equation}
where $\mathbf{Z}(0)$ denotes the value of $\mathbf{Z}$ in the
beginning of time step $h$. The differential operator
$\mathcal{C}=\mathcal{A}+\mathcal{B}$ is approximately expressed
as a series of products of $\mathcal{A}$ and $\mathcal{B}$:
\begin{equation}
\mathcal{C}(h)\approx \Pi^{e}_{j=1}\mathcal{A}(h\alpha_j)
\mathcal{B}(h\beta_j)+O(h^{d+1}),
\end{equation}
where coefficients $\alpha_j$ and $\beta_j$ are determined by the
conditions of order $d$. In this manner, symplectic numerical
integrators of arbitrary orders are built.

If $d=2$, then Equation (3) is the Verlet algorithm (Swope et al.
1982)
\begin{equation}
\mathcal{S}_2(h)=\mathcal{A}(\frac{h}{2})
\mathcal{B}(h)\mathcal{A}(\frac{h}{2}).
\end{equation}
This algorithm is an explicit standard symplectic leapfrog method.
When $d=4$, Equation (3) corresponds to the  explicit symplectic
algorithm of Forest $\&$ Ruth (1990)
\begin{eqnarray}
FR4(h) &=& \mathcal{A}(\frac{\gamma}{2}h) \mathcal{B}(\gamma
h)\mathcal{A}(\frac{1-\gamma}{2}h) \mathcal{B}((1-2\gamma)h)
\nonumber
\\ && \circ
\mathcal{A}(\frac{1-\gamma}{2}h)\mathcal{B}(\gamma
h)\mathcal{A}(\frac{\gamma}{2}h),
\end{eqnarray}
where $\gamma=1/(2-\sqrt[3]{2})$.

Evidently, the construction of these explicit symplectic
integrators is based on the Hamiltonian with an analytically
integrable decomposition. Can such an operator-splitting technique
be available in strictly general relativistic systems, such as a
Schwarzschild space-time? The succeeding discussions answer this
question.

\section{Schwarzschild black holes}

A Schwarzschild black hole with mass $M$ is a nonrotating black
hole. In spherical-like coordinates $(t, r, \theta, \phi)$, the
Schwarzschild metric is described by
\begin{eqnarray}
-c^2d\tau^2 &=& ds^{2} = g_{\alpha\beta}dx^{\alpha}dx^{\beta} \nonumber \\
&=&
-(1-\frac{2GM}{rc^2}) c^2dt^{2} +(1-\frac{2GM}{rc^2})^{-1}   \nonumber \\
& & \cdot dr^{2}+r^{2}d \theta^{2} +r^{2}\sin^{2} \theta d \phi^2,
\end{eqnarray}
where $\tau$, $c$ and $G$ denote proper time, the speed of light
and constant of gravity, respectively. In general, $c$ and $G$ use
geometrized units, $c=G=1$. $M$ also has one unit, $M=1$. This
unit mass can be obtained via scale transformations to certain
quantities: $t\rightarrow tM$, $r\rightarrow rM$ and
$\tau\rightarrow \tau M$. In this manner, this metric is
transformed into a dimensionless form as follows:
\begin{eqnarray}
-d\tau^2 = ds^{2} &=& -(1-\frac{2}{r})dt^{2} +(1-\frac{2}{r})^{-1}
dr^{2}
\nonumber \\
& &+r^{2}d \theta^{2} +r^{2}\sin^{2} \theta d \phi^2.
\end{eqnarray}

This metric corresponds to a Lagrangian system
\begin{equation}
\mathcal{L} = \frac{1}{2} (\frac{ds}{d\tau})^2
=\frac{1}{2}g_{\mu\nu}\dot{x}^{\mu}\dot{x}^{\nu},
\end{equation}
where $\dot{x}^{\mu}=\mathbf{U}$ is a four-velocity. A covariant
generalized momentum $\mathbf{p}$ is defined in the following form
\begin{equation}
p_{\mu} = \frac{\partial \mathcal{L}}{\partial
\dot{x}^{\mu}}=g_{\mu\nu}\dot{x}^{\nu}.
\end{equation}
This Lagrangian does not explicitly depend on $t$ and $\phi$, and
thus, two constant momentum components exist. They are
\begin{eqnarray}
p_{t} &=& -(1-\frac{2}{r})\dot{t}=-E,\\
p_{\phi} &=& r^{2}\sin^{2}\theta\dot{\phi}=\ell,
\end{eqnarray}
where $E$ and $\ell$ are the energy and angular momentum of a test
particle moving around a black hole, respectively.

In accordance with classical mechanics, a Hamiltonian derived from
the Lagrangian is expressed as
\begin{eqnarray}
\mathcal{H} &=&\mathbf{U}\cdot\mathbf{p}-\mathcal{L}
=\frac{1}{2}g^{\mu\nu}p_{\mu}p_{\nu} =
-\frac{1}{2}(1-\frac{2}{r})^{-1} E^{2}
\nonumber \\
&& +\frac{1}{2}(1-\frac{2}{r})p^{2}_{r}
+\frac{1}{2}\frac{p^{2}_{\theta}}{r^2}+\frac{1}{2}\frac{\ell^{2}}{r^2\sin^2\theta}.
\end{eqnarray}
This Hamiltonian governs the motion of a test particle around the
Schwarzschild black hole.

A point is worth noting. A magnetic field arises due to the
relativistic motion of charged particles in an accretion disc
around the central black hole (Borm $\&$ Spaans 2013). It also
leads to generating gigantic jets along the magnetic axes. The
magnetic field is too weak to change the gravitational background
and alter the metric tensor of the Schwarzschild black hole
space-time.  However, it can exert a considerable influence on the
motion of charged test particles. Considering this point, we
suppose that the particle has a charge $q$ and the black hole is
immersed into an external asymptotically uniform magnetic field.
The magnetic field is parallel to the $z$-axis, and its strength
is $B$. The electromagnetic four-vector potential $A^{\alpha}$ in
the Lorentz gauge is a linear combination of the time-like and
space-like axial Killing vectors $\xi^{\alpha}_{(t)}$ and
$\xi^{\alpha}_{(\phi)}$ (Abdujabbarov et al. 2013; Shaymatov et
al. 2015; Tursunov et al. 2016; Benavides-Gallego et al. 2019):
\begin{equation}
A^{\alpha}=C_1\xi^{\alpha}_{(t)}+C_2\xi^{\alpha}_{(\phi)}.
\end{equation}
In  Felice $\&$ Sorge (2003), the constants are set as $C_1=0$ and
$C_2=B/2$. In this manner, the four-vector potential has only one
nonzero covariant component
\begin{equation}
A_{\phi}=\frac{B}{2}g_{\phi\phi}=\frac{B}{2}r^{2}\sin^{2} \theta.
\end{equation}
The charged particle motion is described by the Hamiltonian system
\begin{eqnarray}
K &=& \frac{1}{2}g^{\mu\nu}(p_{\mu}-qA_{\mu})(p_{\nu} -qA_{\nu})
\nonumber \\
&=& -\frac{1}{2}(1-\frac{2}{r})^{-1} E^{2}
+\frac{1}{2}(1-\frac{2}{r})p^{2}_{r}
+\frac{1}{2}\frac{p^{2}_{\theta}}{r^2} \nonumber \\
&& +\frac{1}{2r^2\sin^2\theta}(L-\frac{\beta}{2}r^{2}\sin^{2}
\theta)^{2},
\end{eqnarray}
where $\beta=qB$. The energy $E$ is still determined using
Equation (10). However, the expression of angular momentum is
dissimilar to that of Equation (11) and is presented as
\begin{equation}
L= r^{2}\sin^{2}\theta\dot{\phi}+\frac{\beta}{2}r^{2}\sin^{2}
\theta.
\end{equation}
A point is illustrated here. The dimensionless Hamiltonian (15) is
obtained after scale transformations of $B\rightarrow B/M$,
$E\rightarrow mE$, $p_r\rightarrow mp_r$, $q\rightarrow mq$,
$L\rightarrow mML$, $p_{\theta}\rightarrow mMp_{\theta}$ and
$K\rightarrow m^2K$, where $m$ is the particle's mass. In
addition, the Schwarzschild solution with an external magnetic
field is the Hamiltonian (15), and it no longer has a background
solution to general relativity.

The Hamiltonians $\mathcal{H}$ and $K$ always remain at a given
constant as follows:
\begin{eqnarray}
\mathcal{H} &=& -\frac{1}{2}, \\
K &=& -\frac{1}{2}.
\end{eqnarray}
They are attributed to the four-velocity relation
$\mathbf{U}\cdot\mathbf{U}=-1$. In addition, a second integral
(i.e., the Carter constant) can be easily found in the Hamiltonian
$\mathcal{H}$ by performing the separation of variables in the
Hamilton-Jacobi equation. Thus, this Hamiltonian is integrable and
has formal analytical solutions. However, the perturbation from
the external magnetic field leads to the absence of a second
integral. In such case, no formal analytical solutions exist in
the Hamiltonian $K$.

\section{Construction of explicit symplectic integrators for Schwarzschild space-times}

Suppose the Hamiltonian (12) is similar to the Hamiltonian (1) and
has two splitting parts:
\begin{eqnarray}
\mathcal{H} &=& \mathcal{T}+\mathcal{V}, \\
\mathcal{T} &=& \frac{1}{2}(1-\frac{2}{r})p^{2}_{r}
+\frac{1}{2}\frac{p^{2}_{\theta}}{r^2}, \\
\mathcal{V} &=& -\frac{1}{2}(1-\frac{2}{r})^{-1} E^{2}
+\frac{1}{2}\frac{\ell^{2}}{r^2\sin^2\theta}.
\end{eqnarray}
The $\mathcal{V}$ part is analytically integrable, and its
analytical solutions $p_r$ and $p_{\theta}$ are explicit functions
of proper time $\tau$. Although the $\mathcal{T}$ part exhibits no
separation of variables, it is still analytically integrable.
However, its analytical solutions $r$ and $p_{r}$ are not explicit
functions of proper time $\tau$ but are implicit functions. In
such case, the explicit symplectic integrators in Equations (4)
and (5) are unsuitable for the Hamiltonian splitting form (19).
Consequently, implicit symplectic integrators rather than explicit
ones can be constructed in relativistic Hamiltonian systems, such
as Equation (12), in the general case. The $\mathcal{V}$ part is
more complicated and is not a separation of variables in most
cases in general relativity. Thus, the construction of explicit
symplectic methods becomes more difficult.

From the preceding demonstrations, the key for constructing
explicit symplectic integrators requires the integrated
Hamiltonian to exist as an analytically integrable decomposition.
In particular, the obtained analytical solutions for each
splitting part should be explicit functions of proper time $\tau$.
In summary, the two points must be satisfied for constructing
explicit symplectic integrators. The Hamiltonian (12) with the two
analytically integrable splitting parts fails to construct any
explicit symplectic scheme. Subsequently, we focus on the
Hamiltonian with more analytically integrable splitting parts.

We split the Hamiltonian $\mathcal{H}$ into four pieces:
\begin{equation}
\mathcal{H}=\mathcal{H}_1+\mathcal{H}_2+\mathcal{H}_3+\mathcal{H}_4,
\end{equation}
where these sub-Hamiltonians are
\begin{eqnarray}
\mathcal{H}_1 &=&\frac{1}{2}\frac{\ell^{2}}{r^2\sin^2\theta}
-\frac{1}{2}(1-\frac{2}{r})^{-1} E^{2}, \\
\mathcal{H}_{2} &=& \frac{1}{2}p^{2}_{r},\\
\mathcal{H}_{3} &=& -\frac{1}{r}p^{2}_{r},\\
\mathcal{H}_{4} &=& \frac{p^{2}_{\theta}}{2r^2}.
\end{eqnarray}

For the sub-Hamiltonian $\mathcal{H}_1$, its canonical equations
are $\dot{r}=\dot{\theta}=0$ and
\begin{eqnarray}
\frac{dp_{r}}{d\tau} &=& -\frac{\partial \mathcal{H}_1}{\partial
r}
= \frac{\ell^{2}} {r^3\sin^2\theta}-\frac{E^{2}}{(r-2)^{2}},\\
\frac{dp_{\theta}}{d\tau} &=& -\frac{\partial
\mathcal{H}_1}{\partial \theta}= \frac{\ell^2\cos\theta}
{r^{2}\sin^{3}\theta}.
\end{eqnarray}
Evidently, $r$ and $\theta$ are constants when proper time goes
from $\tau_0$ to $\tau_1=\tau_0+\tau$. Thus, $p_r$ and
$p_{\theta}$ can be solved analytically from Equations (27) and
(28). They are explicit functions of $\tau$ in the following forms
\begin{eqnarray}
p_{r}(\tau) &=& p_{r0} +\tau
[\frac{\ell^{2}}{r^3_0\sin^2\theta_0}-\frac{E^{2}}{(r_0-2)^{2}}],\\
p_{\theta}(\tau) &=& p_{\theta0}+\tau \frac{\ell^2\cos\theta_0}
{r^{2}_0\sin^{3}\theta_0},
\end{eqnarray}
where $r_0$, $\theta_0$, $ p_{r0}$ and $p_{\theta0}$ represent
values of $r$, $\theta$, $ p_{r}$ and $p_{\theta}$ at the proper
time $\tau_0$; and $ p_{r}(\tau)$ and $p_{\theta}(\tau)$ denote
the values of $ p_{r}$ and $p_{\theta}$ at proper time $\tau_1$. A
differential operator for solving $\mathcal{H}_1$ is labeled as
$\psi^{\mathcal{H}_1}_{\tau}$.

The canonical equations of the sub-Hamiltonians $\mathcal{H}_2$,
$\mathcal{H}_3$ and $\mathcal{H}_4$ are
\begin{eqnarray}
\mathcal{H}_2: ~ \frac{dr}{d\tau} &=& p_{r}, ~~\dot{p}_{r}=0; \\
\mathcal{H}_3: ~ \frac{dr}{d\tau} &=& -\frac{2}{r}p_r, ~
\frac{d p_r}{d\tau} = -\frac{p^{2}_r}{r^2}; \\
\mathcal{H}_4: ~ \frac{d\theta}{d\tau} &=&
\frac{p_{\theta}}{r^{2}}, ~ \frac{dp_r}{d\tau} =
\frac{p^2_{\theta}}{r^{3}}, ~ \dot{r}=\dot{p}_{\theta}=0.
\end{eqnarray}
Let $\psi^{\mathcal{H}_2}_{\tau}$, $\psi^{\mathcal{H}_3}_{\tau}$
and $\psi^{\mathcal{H}_4}_{\tau}$ be three operators. We obtain
the solutions for Equations (31)-(33) as follows:
\begin{eqnarray}
\psi^{\mathcal{H}_2}_{\tau}: ~r(\tau) &=& r_0+\tau p_{r0}; \\
\psi^{\mathcal{H}_3}_{\tau}: ~ r(\tau) &=& [(r^{2}_{0}-3\tau
p_{r0})^{2}/r_0]^{1/3}, \nonumber \\
 p_r(\tau) &=& p_{r0}[(r^{2}_{0}-3\tau
p_{r0})/r^2_0]^{1/3}; \\
\psi^{\mathcal{H}_4}_{\tau}: ~ \theta(\tau) &=& \theta_0+\tau p_{\theta0}/r^{2}_{0}, \nonumber \\
 p_r(\tau) &=& p_{r0}+\tau p^2_{\theta0}/r^{3}_{0}.
\end{eqnarray}
It is clear that these solutions are explicit functions of proper
time $\tau$. If the sum of $\mathcal{H}_2$ and $\mathcal{H}_3$ is
regarded as an independent sub-Hamiltonian, then it is
analytically solved. However, the analytical solutions of $r$,
$\theta$ and $p_r$ for the sum cannot be expressed as explicit
functions of proper time $\tau$. Thus, such a composed
sub-Hamiltonian is not considered. Equation (22) is a possible
Hamiltonian splitting for satisfying this requirement. Other
appropriate splitting forms may be provided to the Hamiltonian
(12).

The flow  $\psi^{\mathcal{H}}_{h}$ of the Hamiltonian (12) over
time step $h$ is approximately given by the symmetric composition
of these operators
\begin{eqnarray}
\psi^{\mathcal{H}}_{h}\approx S^{\mathcal{H}}_2(h) &=&
\psi^{\mathcal{H}_4}_{h/2}\circ \psi^{\mathcal{H}_3}_{h/2}\circ
\psi^{\mathcal{H}_2}_{h/2}\circ \psi^{\mathcal{H}_1}_{h} \nonumber
\\ & & \circ \psi^{\mathcal{H}_2}_{h/2}\circ \psi^{\mathcal{H}_3}_{h/2}\circ
\psi^{\mathcal{H}_4}_{h/2}.
\end{eqnarray}
The above construction is a second order explicit symplectic
integrator marked as $S^{\mathcal{H}}_2$. Its difference scheme is
provided in Appendix A.

The order of algorithm (37) can be lifted to four by using the
composition scheme of Yoshida (1990). That is, a fourth order
symplectic composition construction is
\begin{equation}
S^{\mathcal{H}}_4(h)=S^{\mathcal{H}}_2(\gamma h)\circ
S^{\mathcal{H}}_2(\delta h)\circ S^{\mathcal{H}}_2(\gamma h),
\end{equation}
where $\delta=1-2\gamma$.

The Hamiltonian (15) exhibits the following splitting form
\begin{equation}
K=K_1+K_2+K_3+K_4,
\end{equation}
where $K_2=\mathcal{H}_2$, $K_3=\mathcal{H}_3$,
$K_4=\mathcal{H}_4$, and the inclusion of $A_{\phi}$ only changes
$\mathcal{H}_1$ as
\begin{eqnarray}
K_{1} &=& \frac{1}{2r^2\sin^2\theta}(L-\frac{\beta}{2}
r^{2}\sin^{2} \theta)^{2} \nonumber \\ &&
-\frac{1}{2}(1-\frac{2}{r})^{-1} E^{2}.
\end{eqnarray}
When $\mathcal{H}_1$ gives place to $K_1$, the explicit symplectic
integrators $S_2$ and $S_4$ are still suitable for the
non-integrable Hamiltonian $K$ of the Schwarzschild solution with
an external magnetic field, labeled as $S^{K}_2$ and $S^{K}_4$.

In summary, when the Hamiltonians (12) and (15) are split into
four analytically integrable parts, their explicit symplectic
integrators are easily constructed.

\begin{table*}[htbp]
\centering \caption{Dependence of stable (S) or unstable (U)
behavior of Hamiltonian errors for the seven algorithms on step
size $h$. Chaotic Orbit 3 in Figure 2 is integrated until proper
time $\tau=10^{8}$.} \label{Tab1}
\begin{tabular}{lccccccccc}
\hline Method   & S2   & EI2 &  EE2  & S4  & EI4  & EE4  & RK4\\
\hline $h=0.1$  & S & S & S & U & U & S & U \\
\hline $h=1.0$ & S & S & U & S & U & S & U\\
\hline $h=10$  & S & S & U & S & S & U & U \\
\hline
\end{tabular}
\end{table*}

\begin{table*}[htbp]
\centering \caption{Same as Table 1, but dependence of the largest
absolute values of Hamiltonian errors on $h$.} \label{Tab2}
\begin{tabular}{lccccccccc}
\hline Method  & S2   & EI2 &  EE2  & S4  & EI4  & EE4  & RK4\\
\hline $h=0.1$ & 4e-8 & 4e-8 & 3e-8 & 7e-9  & 3e-12 & 1e-12 & 4e-12 \\
\hline $h=1.0$ & 6e-6 & 5e-6 & 2e-6 & 3e-8 & 7e-9 & 2e-8 & 4e-7\\
\hline $h=10$  & 8e-4 & 6e-3 & 6e-3 & 4e-4 & 7e-5 & 4e-3 & 3e-2 \\
\hline
\end{tabular}
\end{table*}

\begin{table*}[htbp]
\centering \caption{Same as Table 1, but dependence of
computational cost, i.e., CPU times (minute: second), on $h$.}
\label{Tab3}
\begin{tabular}{lccccccccc}
\hline Method     & S2   & EI2   &  EE2  & S4    & EI4   & EE4   & RK4\\
\hline $h=0.1$    & 9:13 & 10:13 & 14:22 & 27:42 & 30:33 & 33:35 & 17:48\\
\hline $h=1.0$    & 0:56 & 1:03  & 1:26  & 2:46  & 3:09  & 3:21  & 1:46\\
\hline $h=10$     & 0:05 & 0:07  & 0:07  & 0:16  & 0:20  & 0:19  & 0:10 \\
\hline
\end{tabular}
\end{table*}

\section{Numerical evaluations}

In this section, we focus on checking the numerical performance of
the proposed integrators. For comparison, a conventional
fourth-order Runge-Kutta integrator (RK4), second- and
fourth-order symplectic algorithms consisting of explicit and
implicit mixed methods (EI2 and EI4), and second- and fourth-order
extended phase-space explicit symplectic-like methods (EE2 and
EE4) are used. The details of EI2, EI4, EE2 and EE4 are provided
in Appendix B.

\subsection{Case of $\beta=0$}

When no charges are assigned to test particles, the system (15) is
transformed to the Schwarzschild problem (12). We consider
parameters $E=0.995$ and $\ell$ (or $L$) =4.6, and  proper time
step size $h=1$. Initial conditions are $r=11$, $\theta=\pi/2$ and
$p_r=0$. The initial value of $p_{\theta}$ ($>0$) is determined by
using Equation (17). We conduct our numerical experiments by
applying each of the aforementioned algorithms to solve the
Hamiltonian (12). As shown in Figure 1(a), the three second-order
methods, namely, S2, EI2 and EE2, provide an order of $10^{-6}$ to
Hamiltonian errors $\Delta H=1+2\mathcal{H}$ from Equation (17) at
the end of integration time. Differences exist among the
algorithmic errors. The new symplectic algorithm S2 and the
explicit and implicit mixed symplectic method EI2 have nearly the
same errors, which remain bounded and stable. This result
indicates the superiority of S2 in the conservation of the
long-term stable behavior of energy (or Hamiltonian) errors.
However, the extended phase-space method EE2 exhibits an increase
in secular errors. This increase can be prevented if a small time
size $h=0.1$ is used. In such case, the errors (not plotted) can
be stabilized within an order of $10^{-8}$.

The four fourth-order algorithms, namely,  S4, EI4, EE4 and RK4,
yield the Hamiltonian errors in Figures 1(b) and 1(c). The
algorithms S4, EI4 and EE4 are accurate to an order of $10^{-8}$.
The new method S4 and the extended phase-space method EE4 have
stable and bounded errors. The explicit and implicit mixed
symplectic method EI4 causes the errors to become bounded.
Meanwhile, RK4 provides the lowest accuracy with an order of
$10^{-6}$ and its errors increase linearly with time. This result
is expected because RK4 is not a geometric integrator.

The considered orbit, called Orbit 1, can be observed from the
Poincar\'{e} section map on the plane $\theta=\pi/2$ and
$p_{\theta}>0$. The map relates to a two-dimensional plane, which
exhibits intersections of the particles' trajectories with the
surface of section in phase space (Lichtenberg $\&$ Lieberman
1983). If the plotted points form a closed curve, then the motion
is regular. This result is based on  a regular trajectory moving
on a torus in the phase space and the curve being a cross section
of the torus. By contrast, if the plotted points are distributed
randomly, then the motion is chaotic. With the aid of the
distribution of the points in the Poincar\'{e} map, we can
determine the phase-space structure, indicating whether the motion
is chaotic. The Kolmogorov-Arnold-Moser (KAM) torus in the section
in Figure 1(d) is provided by the new method S2 and indicates the
regularity of Orbit 1. In addition, the structure of Orbit 1, and
those of Orbits 2 and 3 with initial separations $r=70$ and 110
are described, respectively. The numerical performance of the
aforementioned algorithms acting on Orbit 1 is approximately
consistent with those acting on Orbits 2 and 3.

\subsection{Case of $\beta\neq 0$}

When an external magnetic field with parameter $\beta=8.9\times
10^{-4}$ is included within the vicinity of a black hole, the
system is non-integrable. The magnetic field causes the three
orbits in Figure 1(d) to have different phase-space structures in
Figure 2(a). Although Orbit 1 remains a simply closed torus, it is
shrunk drastically and becomes a small torus. By contrast, Orbit 2
becomes a more complicated KAM torus, consisting of seven small
loops wherein the successive points jump from one loop to the
next. These small loops  belong  to the same trajectory and form a
chain of islands (H\'{e}non $\&$ Heiles 1964). Such a torus is
regular but easily induces the occurrence of resonance and chaos.
In particular, Orbit 3, which is a small loop in Figure 1(d), is
considerably enlarged and densely filled in the phase space. This
result indicates the onset of strong chaoticity.

Although the loop of Orbit 1 is considerably smaller under the
interaction of the electromagnetic forces in Figure 2(a) than in
the case without electromagnetic forces in Figure 1(d), each
algorithm exhibits nearly the same performance in the two cases
because the tori of Orbit 1 in the two cases belong to the same
category of trajectories, namely, simple single regular loops.
Orbits 2 and 3 exhibit completely different dynamical behavior,
but correspond to approximately the same Hamiltonian errors for
each integration method. Figures 2(b)-2(d) plot the errors for
chaotic Orbit 3. The errors of the second-order methods for
chaotic Orbit 3 shown in Figure 2(b) are approximately consistent
with those for regular Orbit 1 shown in Figure 1(a). The
fourth-order algorithms S4 and EE4 exhibit no dramatic differences
in errors in Figure 2(c), similar to that in Figure 1(b). This
result indicates that orbital chaoticity does not explicitly
affect algorithmic accuracy. However, the explicit and implicit
mixed method EI4 presents a secular drift in errors due to
roundoff errors. The increase in errors can be prevented when a
large time size $h=10$ is adopted. In such case, accuracy is
maintained with an order of $10^{-5}$. EI4 exhibits secular drift
in the Hamiltonian errors for the smaller time step $h=1$ but does
not for the larger time size $h=10$. The following is a simple
analysis. The errors of a symplectic integrator mostly consist of
truncation and roundoff errors. When truncation errors are more
than roundoff errors, the symplectic integrator causes the
Hamiltonian errors to remain bounded and to exhibit no secular
drift in appropriate situations. Roundoff errors increase with an
increase in the number $N$ of calculations. They are approximately
estimated using $N\epsilon$, where $\epsilon\sim 10^{-16}$
demonstrates machine precision in double floating-point precision.
When roundoff errors completely dominate total errors, the
Hamiltonian or energy errors increase linearly with time. Assume
that a symplectic method has a truncation energy error in an order
of $10^{-12}$. The  total errors in the energy are stabilized at
the order of magnitude when $N<10^{4}$, but grow linearly as
$N\gg10^{4}$. If a symplectic method has a truncation energy error
higher than the order of $10^{-8}$, then the  total  errors in the
energy remain bounded and approach the order of truncation errors
when $N<10^{8}$, whereas increase linearly as $N\gg10^{8}$. These
results have been confirmed by numerical experiments on $N$-body
problems in the solar system (Wu et al. 2003; Deng et al. 2020).
In the present numerical simulations, the truncation Hamiltonian
errors of EI4 are in the order of $10^{-9}$ for $h=1$ but the
roundoff errors are $10^{-8}$ after $10^{8}$ integration steps.
Given that the former errors are smaller than the latter ones,
secular drift exists in the Hamiltonian errors. However, the
truncation Hamiltonian errors of EI4 are in the order of $10^{-5}$
for $h=10$. They are larger than  the roundoff errors after
$10^{8}$ integration steps. Therefore, no secular drift occurs in
the Hamiltonian errors.

A conclusion can be drawn from  Figures 1 and 2 that the stable
behavior and magnitude of the Hamiltonian errors for each
algorithm mostly depend on the choice of step sizes. To
demonstrate this fact clearly, we list them in Tables 1 and 2,
where chaotic Orbit 3 is used as a test orbit. The two
second-order symplectic integrators S2 and EI2 can make the errors
bounded for the three time steps, $h=0.1, 1, 10$. A larger time
step is also suitable for the two fourth-order symplectic
integrators S4 and EI4. However, a smaller time step is suitable
for the extended phase-space methods. The reason why EE2 does not
produce stable errors for $h=1$ but does for $h=0.1$ (or EE4 does
not produce stable errors for $h=10$ but does for $h=1$) differs
from why S4 does not provide stable errors for $h=0.1$ but does
for $h=1$. The error stability or instability for the former case
is mostly dependent on permutations, which are frequently required
in appropriately small times. However, it is primarily related to
the roundoff errors for the latter case. Such a smaller time step
is also necessary for RK4 to obtain higher accuracy, although RK4
does not remain at a stable or bounded value of energy errors.

Computational costs are listed in Table 3. Given the smaller step
sizes, several differences among CPU times exist for the same
order methods. The proposed explicit symplectic integrators
achieve the best computational efficiency compared with the other
algorithms at the same order and time step. The explicit and
implicit mixed symplectic methods require smaller additional
computational labor than the same-order new integrators because
only the solutions of $r$ and $p_r$ in IM2 of Equation (B.2)
should be iterated. Such partially implicit constructions are
faster to compute than the completely implicit integrators.

\section{Conclusions}

The major contribution of this study is the successful
construction of explicit symplectic integration algorithms in
general relativistic Schwarzschild type space-time geometries. The
construction is based on an appropriate splitting form of the
Hamiltonian corresponding to this space-time. The Hamiltonian
exists four integrable separable parts with analytical solutions
as explicit functions of proper time. The solutions from the four
parts are symmetrically composed of second- and fourth-order
explicit symplectic integrators, similar to the standard explicit
symplectic leapfrog methods that split the considered Hamiltonian
into two integrable parts with analytical solutions as explicit
functions of time. The proposed algorithms are still valid for an
external magnetic field included within the vicinity of the black
hole.

Numerical tests show that the newly proposed integration schemes
effectively control Hamiltonian errors without secular changes
when appropriate step sizes are adopted. They are well-behaved in
the simulation of the long-term evolution of regular orbits with
single or many loops and weakly or strongly chaotic orbits.
Appropriately larger step sizes are acceptable for such explicit
symplectic integrators to maintain stable or bounded energy (or
Hamiltonian) errors. Explicit constructions are generally superior
to same order implicit methods in computational efficiency.

In summary, the new methods achieve long-time performance.
Therefore, they are highly appropriate for the long-term numerical
simulations of regular and chaotic motions of charged particles in
the present non-integrable magnetized Schwarzschild space-time
background (Felice $\&$ Sorge 2003; Kolo\v{s} et al. 2015; Yi $\&$
Wu 2020). The methods are also useful for studying the chaotic
motion of a charged particle in a tokamak magnetic field (Cambon
et al. 2014). They are suitable for investigating the capture
cross section of magnetized particles and the magnetized
particles' acceleration mechanism near a black hole with an
external magnetic field (Abdujabbarov et al. 2014). These methods
are applicable to the simulation of the dynamics of charged
particles around a regular black hole with a nonlinear
electromagnetic source (Jawad et al. 2016). Such class of explicit
symplectic integration algorithms will be developed to address
other black hole gravitational problems, such as the
Reissner-Nordstr\"{o}m space-time.

\appendix
\section*{APPENDIX}

\section{Discrete difference scheme of algorithm $S^{\mathcal{H}}_2$}

From an $(n-1)$th step to an $n$th step, algorithm
$S^{\mathcal{H}}_2$ has the following discrete difference scheme:
\begin{eqnarray}
\theta^{\mathcal{H}4} &=& \theta_{n-1}+\frac{h}{2} p_{\theta, n-1}/r^{2}_{n-1}, \nonumber \\
 p^{\mathcal{H}4}_r &=& p_{r,n-1}+\frac{h}{2} p^2_{\theta,n-1}/r^{3}_{n-1}; \nonumber \\
r^{\mathcal{H}3} &=& [(r^{2}_{n-1}-\frac{3}{2}h p^{\mathcal{H}4}_r)^{2}/r_{n-1}]^{1/3}, \nonumber \\
 p^{\mathcal{H}3}_r &=& p^{\mathcal{H}4}_r[(r^{2}_{n-1}-\frac{3}{2}h
p^{\mathcal{H}4}_r)/r^2_{n-1}]^{1/3}; \nonumber \\
r^{\mathcal{H}2} &=& r^{\mathcal{H}3}+\frac{h}{2} p^{\mathcal{H}3}_r; \nonumber \\
p^{\mathcal{H}1}_{r} &=& p^{\mathcal{H}3}_r +h
[\frac{\ell^{2}}{(r^{\mathcal{H}2})^3\sin^2\theta^{\mathcal{H}4}}-\frac{E^{2}}{(r^{\mathcal{H}2}-2)^{2}}], \nonumber \\
p_{\theta n} &=& p_{\theta,n-1}+h
\frac{\ell^2\cos\theta^{\mathcal{H}4}}
{(r^{\mathcal{H}2})^2\sin^{3}\theta^{\mathcal{H}4}}; \nonumber \\
r^{*\mathcal{H}2} &=& r^{\mathcal{H}2}+\frac{h}{2} p^{\mathcal{H}1}_r; \nonumber \\
r_{n} &=& [((r^{*\mathcal{H}2})^2-\frac{3}{2}h p^{\mathcal{H}1}_r)^{2}/r^{*\mathcal{H}2}]^{1/3}, \nonumber \\
p^{*\mathcal{H}3}_r &=&
p^{\mathcal{H}1}_r[((r^{*\mathcal{H}2})^{2}-\frac{3}{2}h
p^{\mathcal{H}1}_r)/(r^{*\mathcal{H}2})^2]^{1/3}; \nonumber \\
\theta_{n} &=& \theta^{\mathcal{H}4}+\frac{h}{2} p_{\theta n}/(r_n)^2, \nonumber \\
 p_{rn} &=& p^{*\mathcal{H}3}_{r}+\frac{h}{2} (p_{\theta n})^2/(r_n)^3. \nonumber
\end{eqnarray}
In this manner, the solutions $(r_{n}, \theta_{n},  p_{rn},
p_{\theta n})$ at the $n$th step are presented. Let the
integration continue from the $n$th step to the  $(n+1)$th step.

\section{Descriptions of algorithms EI4 and EE4}

Algorithm EI4 was discussed in the references (Lubich et al. 2010;
Zhong et al. 2010; Mei et al. 2013a, 2013b). Here, it is used to
solve the Hamiltonian (15). Its construction requires splitting
this Hamiltonian into two parts
\begin{equation}
K=K_1+\Lambda,
\end{equation}
where $\Lambda=K_2+K_3+K_4$. The sub-Hamiltonian $K_1$ does not
depend on momenta $p_r$ and $p_{\theta}$, and thus, it is easily,
explicitly and analytically solved, and then labeled as operator
$\psi^{K_1}_{h}$. Another sub-Hamiltonian $\Lambda$ exhibits
difficulty in providing explicit analytical solutions, but can be
integrated using the second-order implicit midpoint rule (Feng
1986), labeled as operator $IM2(h)$. Similar to the explicit
algorithm $S_2$ in Equation (4), a second-order explicit and
implicit mixed symplectic integrator is symmetrically composed of
two explicit and implicit operators by
\begin{equation}
EI2(h)=\psi^{K_1}_{h/2}\circ IM2(h)\circ\psi^{K_1}_{h/2}.
\end{equation}
Such a mixed symplectic method demonstrates an explicit advantage
over the implicit midpoint method acting on the complete
Hamiltonian $K$ in terms of computational efficiency. The
four-order explicit and implicit mixed symplectic integrator EI4
can be obtained by substituting EI2 into $S^{\mathcal{H}}_2$ in
Equation (38).

Algorithm EE4 is based on the idea of Pihajoki (2015). Its
construction relies on extending the four-dimensional phase-space
variables $(r,\theta, p_r, p_{\theta})$ of the Hamiltonian $K$  to
eight-dimensional phase-space variables
$(r,\theta,\tilde{r},\tilde{\theta}, p_r,$
$p_{\theta},\tilde{p}_r, \tilde{p}_{\theta})$ of a new
Hamiltonian, i.e.,
\begin{equation}
\Gamma=\kappa_1(r,\theta, \tilde{p}_r,
\tilde{p}_{\theta})+\kappa_2(\tilde{r},\tilde{\theta}, p_r,
p_{\theta}),
\end{equation}
where $\kappa_1(r,\theta, \tilde{p}_r,
\tilde{p}_{\theta})=\kappa_2(\tilde{r},\tilde{\theta}, p_r,
p_{\theta})= K(r,\theta, p_r, p_{\theta})$. Evidently, the two
sub-Hamiltonians $\kappa_1$ and $\kappa_2$ are independently,
explicitly and analytically solved, and then labeled as operators
$\psi^{\kappa_1}_{h}$ and $\psi^{\kappa_2}_{h}$. The two operators
are used to yield the second-order symplectic method
$\mathcal{S}_2$ and the Forest-Ruth fourth-order  algorithm FR4,
which are respectively given by Equations (4) and (5) but
$\mathcal{A}$ and $\mathcal{B}$ are respectively replaced with
$\psi^{\kappa_1}$ and $\psi^{\kappa_2}$.

If the two independent Hamiltonians $\kappa_1$ and $\kappa_2$ have
the same initial conditions, then they should have the same
solutions, i.e., $r=\tilde{r}$, $\theta=\tilde{\theta}$,
$\tilde{p}_r= p_r$ and $\tilde{p}_{\theta}=p_{\theta}$. However,
these solutions are not equal because of their couplings in the
methods $\mathcal{S}_2$ and FR4. To make them equal, Pihajoki
(2015), Liu et al. (2016), Luo et al. (2017), Liu et al. (2017),
Luo $\&$ Wu (2017), Li $\&$ Wu (2017) and Wu $\&$ Wu (2018)
introduced permutations between the original variables and their
corresponding extended variables after the implementation of
$\mathcal{S}_2$ or FR4. A good choice is the midpoint permutation
method(Luo et al. 2017):
\begin{eqnarray}
\mathcal{M}: \frac{r+\tilde{r}}{2} &\rightarrow& r =\tilde{r},
~~~~~~~~ \frac{\theta+\tilde{\theta}}{2}\rightarrow
\theta=\tilde{\theta};
\nonumber \\
\frac{p_r+\tilde{p}_r}{2} &\rightarrow& p_r=\tilde{p}_r, ~~
\frac{p_{\theta}+\tilde{p}_{\theta}}{2}\rightarrow p_{\theta}=
\tilde{p}_{\theta}.
\end{eqnarray}
By adding the midpoint permutation map $\mathcal{M}$ after
$\mathcal{S}_2$ or FR4, Luo et al. (2017) obtained algorithms EE2
and EE4 as follows:
\begin{equation}
EE2= \mathcal{M}\otimes \mathcal{S}_2, ~~ EE4= \mathcal{M}\otimes
FR4.
\end{equation}
The inclusion of   $\mathcal{M}$  destroys the symplecticity of
$\mathcal{S}_2$ and  FR4, but EE2 and EE4, similar to the
symplectic schemes $\mathcal{S}_2$ and FR4, still exhibit good
long-term stable behavior in energy errors because of their
symmetry. Thus, they are called explicit symplectic-like
algorithms for the newly extended phase-space Hamiltonian
$\Gamma$.

\section*{Acknowledgments}

The authors are very grateful to a referee for useful suggestions.
This research has been supported by the National Natural Science
Foundation of China [Grant Nos. 11533004, 11973020 (C0035736),
11803020, 41807437, U2031145]
and the Natural Science Foundation of Guangxi (Grant Nos.
2018GXNSFGA281007 and 2019JJD110006).

\newpage

\begin{figure*}[ptb]
\center{
\includegraphics[scale=0.25]{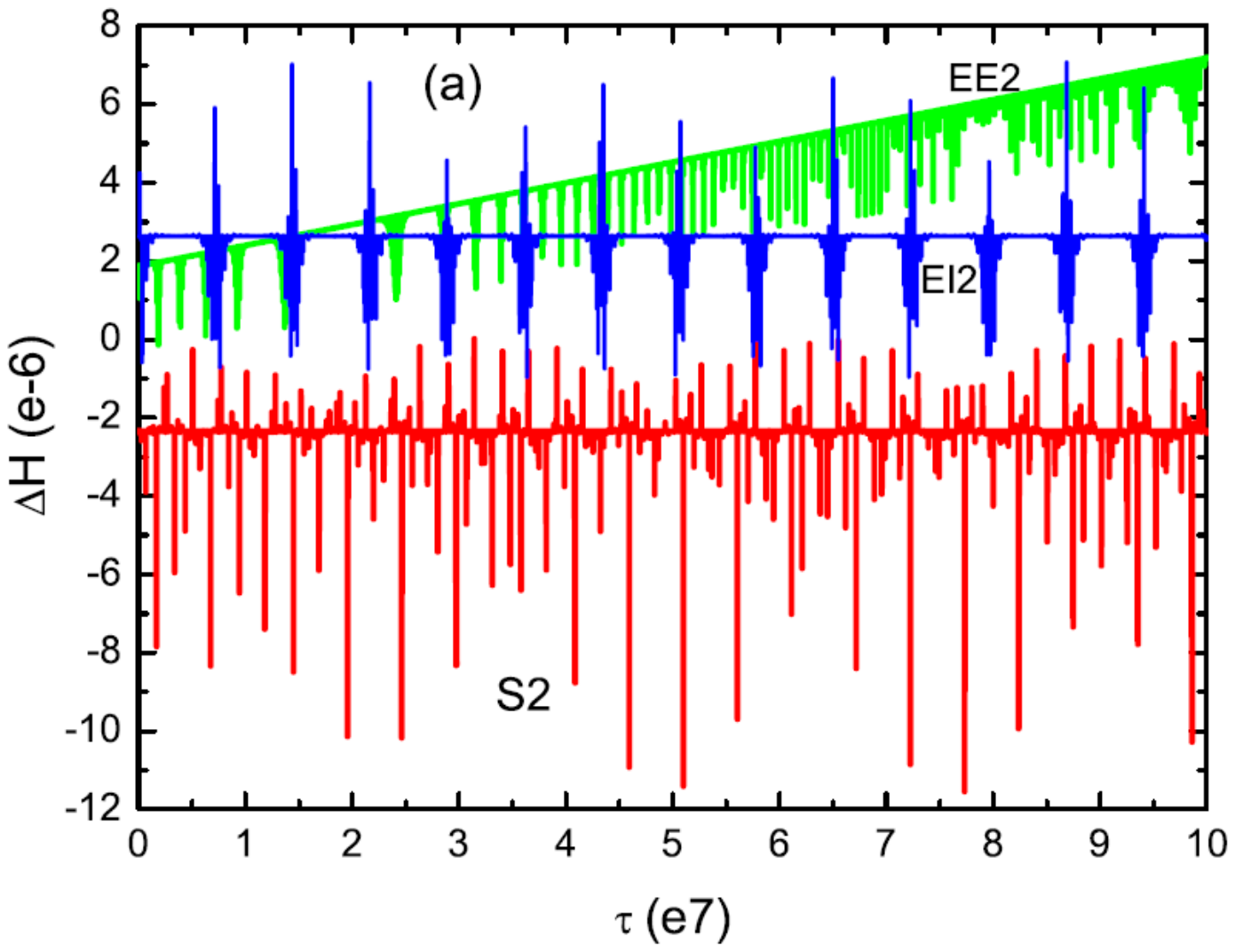}
\includegraphics[scale=0.25]{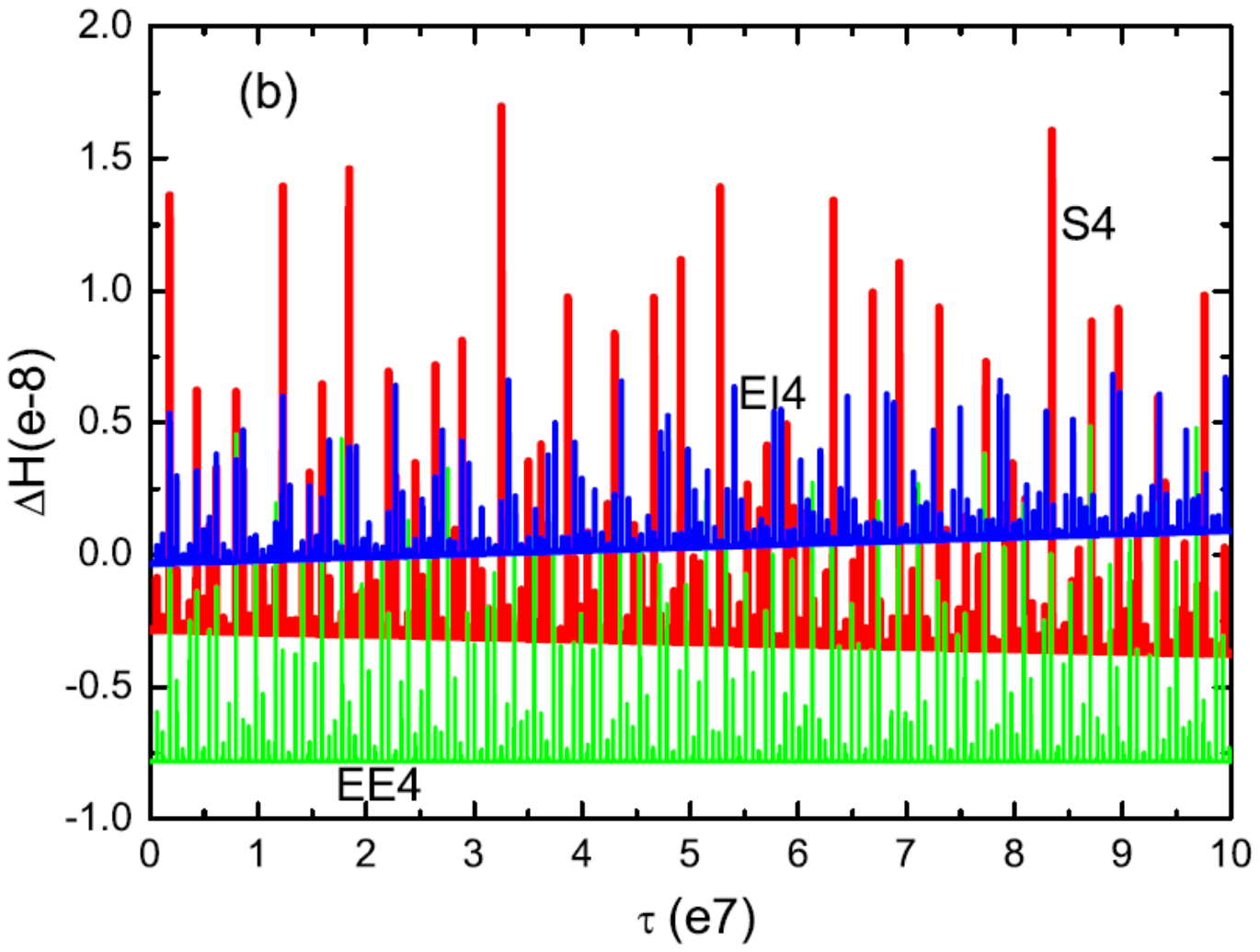}
\includegraphics[scale=0.25]{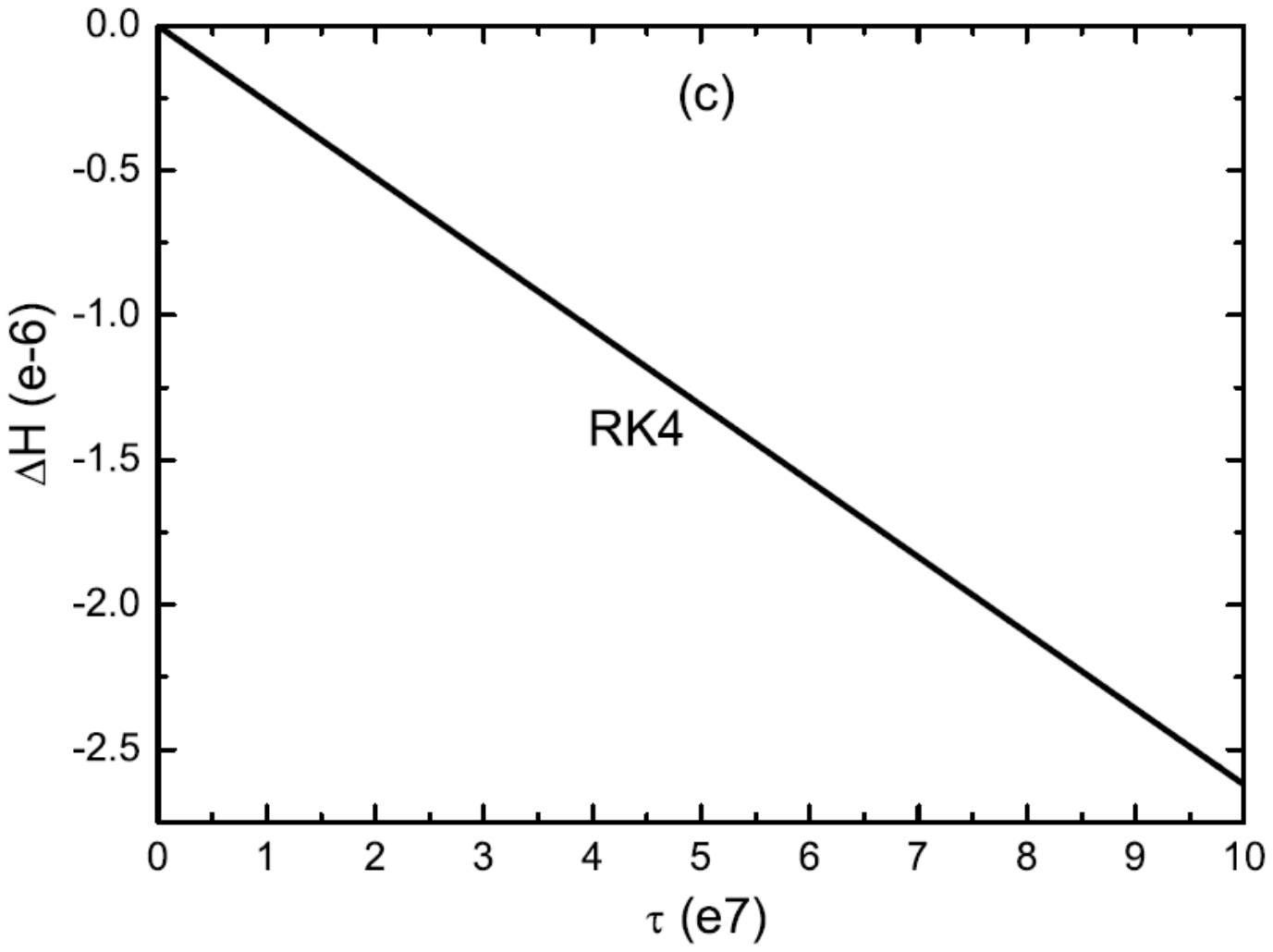}
\includegraphics[scale=0.25]{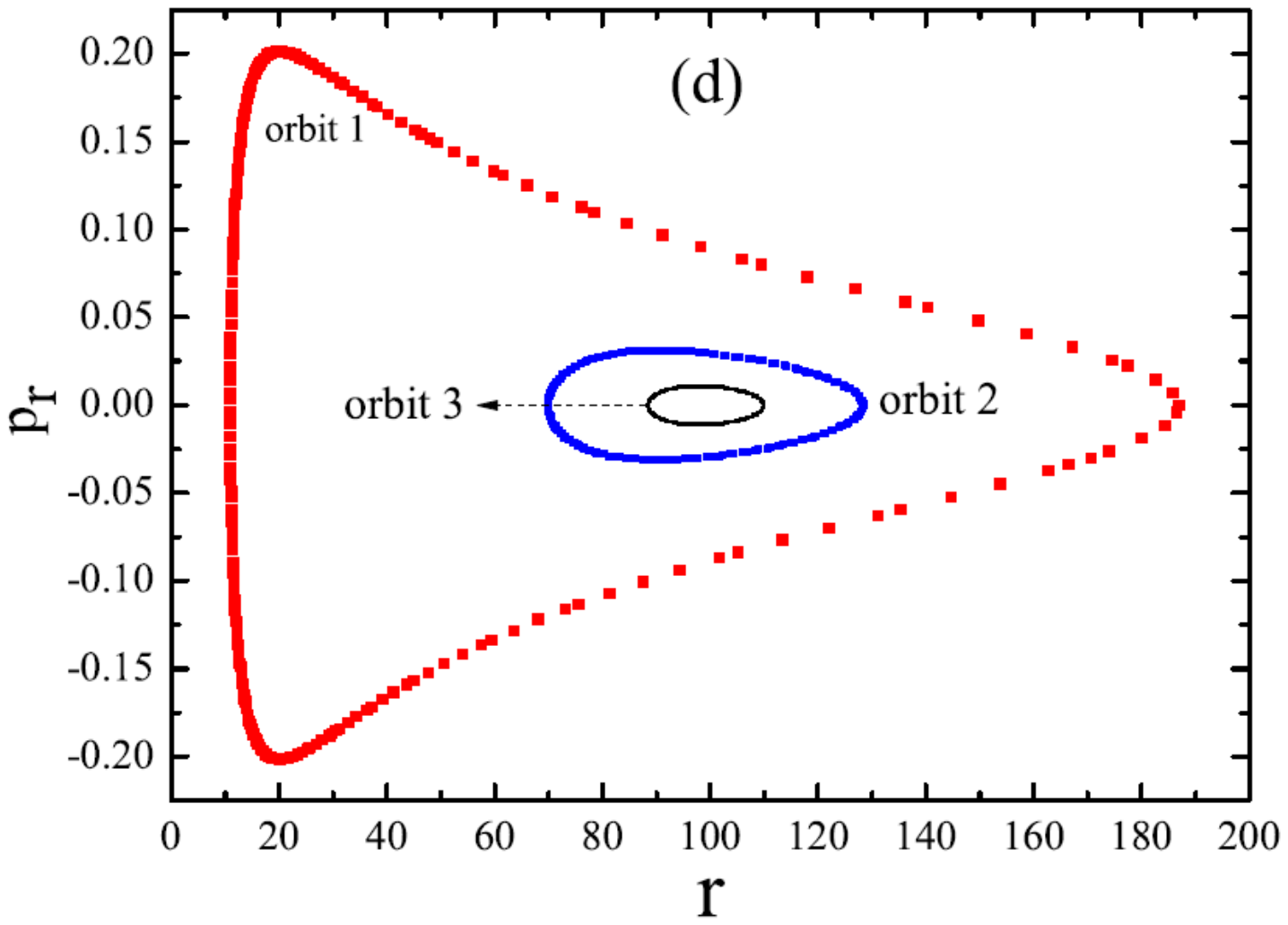}
\caption{(a)-(c) Hamiltonian errors $\Delta H=1+2\mathcal{H}$ from
Eq. (17) for several algorithms solving the Schwarzschild problem
(12). The adopted algorithms are the new second-order explicit
symplectic integrator S2 in Equation (37), the second-order
explicit and implicit mixed symplectic method EI2 in Equation
(B.2), the second-order explicit extended phase-space
symplectic-like algorithm EE2, the new fourth-order explicit
symplectic integrator $S_4$ in Equation (38), the fourth-order
explicit and implicit mixed symplectic method EI4, the
fourth-order explicit extended phase-space symplectic-like
algorithm EE4 in Equation (B.5), and the fourth-order Runge-Kutta
scheme RK4. The energy and angular momentum of particles are
$E=0.995$ and $\ell$ (or $L)$=4.6, respectively, and the proper
time-step is $h=1$. The integrated orbit (called Orbit 1) has
initial conditions $r=11$, $\theta=\pi/2$ and $p_r=0$. The initial
value of $p_{\theta}$ $(>0)$ is given by Equation (17). (d)
Poincar\'{e} sections on the plane $\theta=\pi/2$ and
$p_{\theta}>0$. Apart from Orbit 1, Orbits 2 and 3 with initial
separations $r=70$ and 110, respectively, are plotted. The initial
values of $\theta$ and $p_r$ for Orbits 2 and 3 are the same as
those for Orbit 1. The three orbits are regular tori because of
the integrability of the system (12).}
 \label{Fig1}}
\end{figure*}

\begin{figure*}[ptb]
\center{
\includegraphics[scale=0.25]{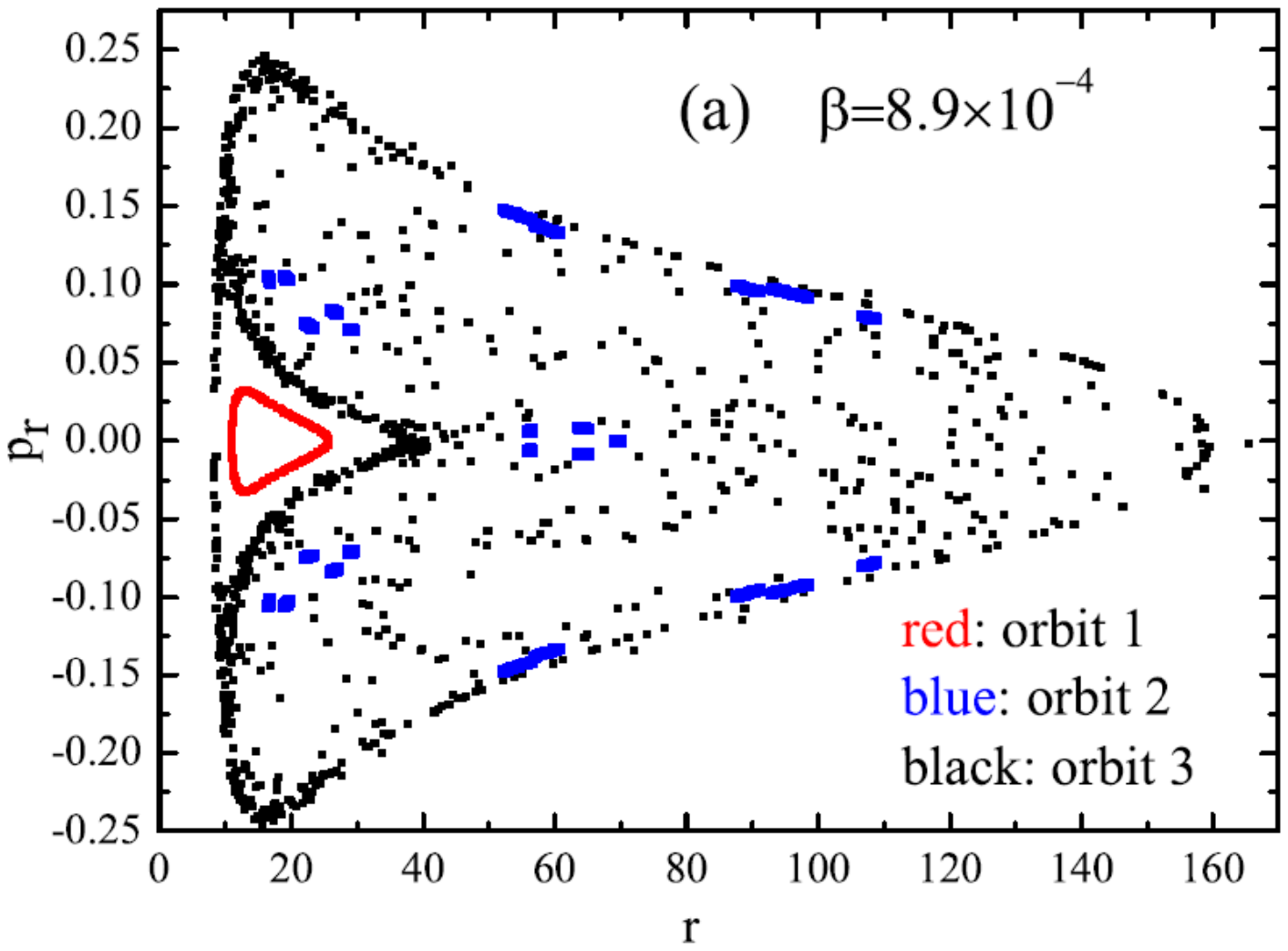}
\includegraphics[scale=0.25]{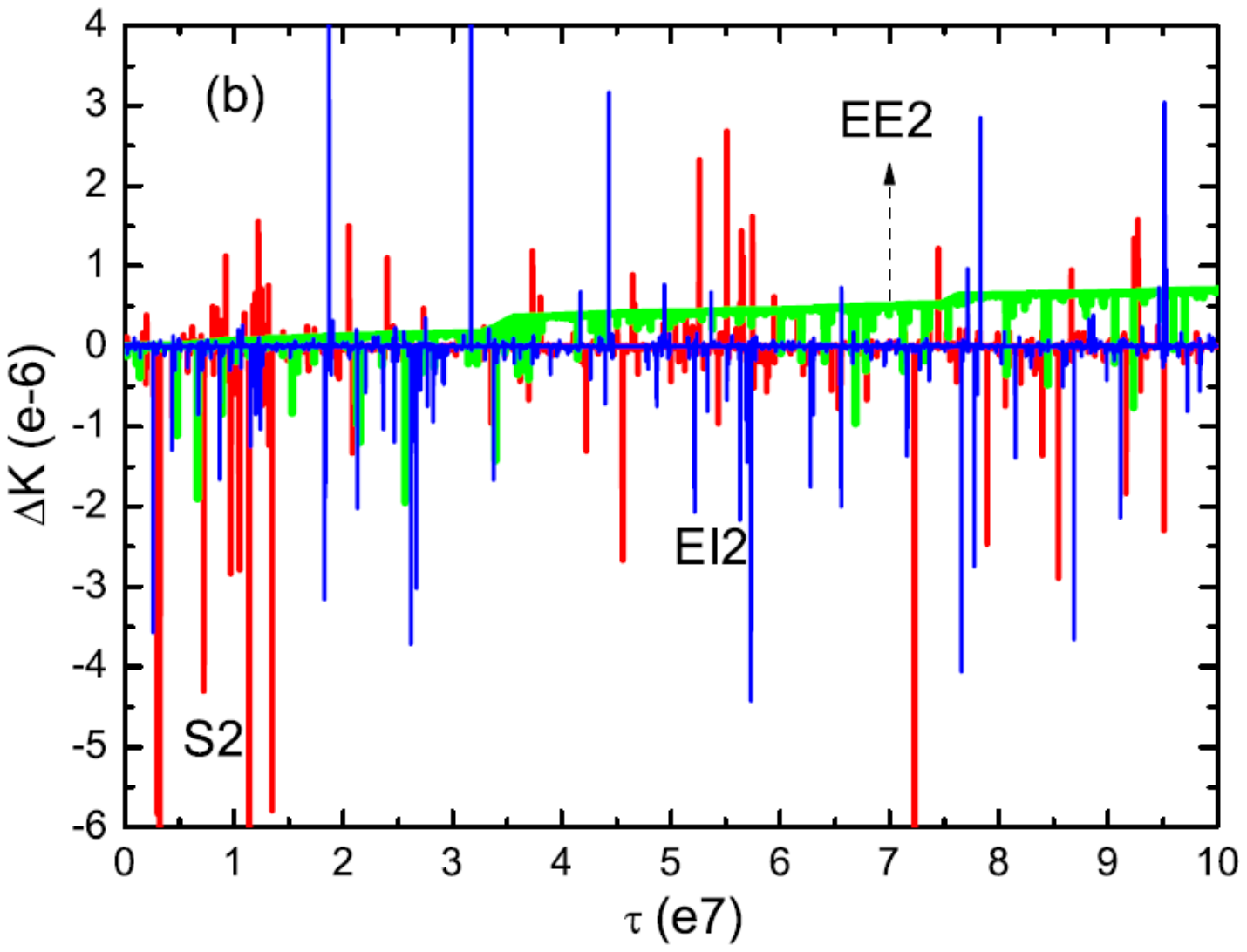}
\includegraphics[scale=0.25]{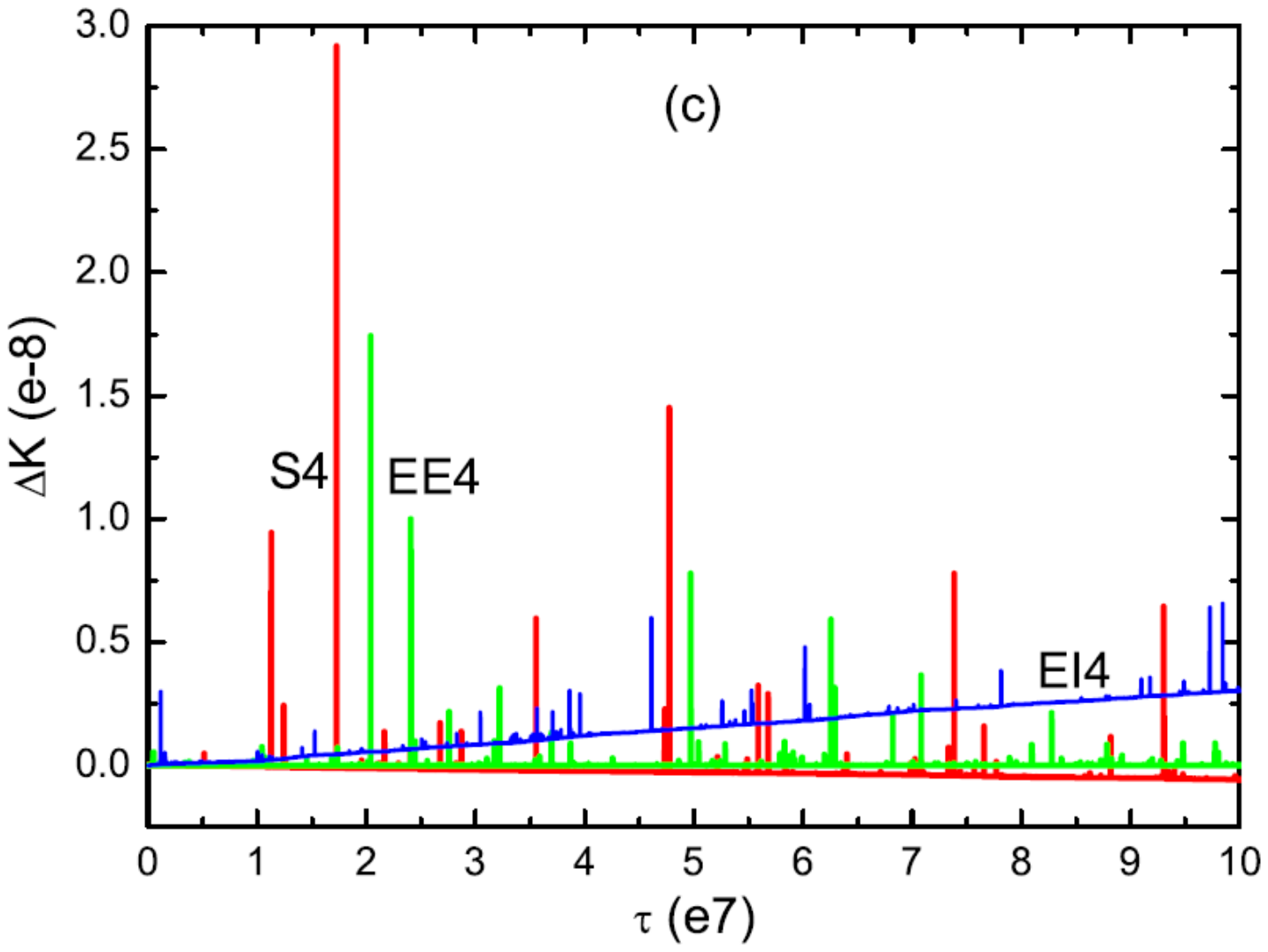}
\includegraphics[scale=0.25]{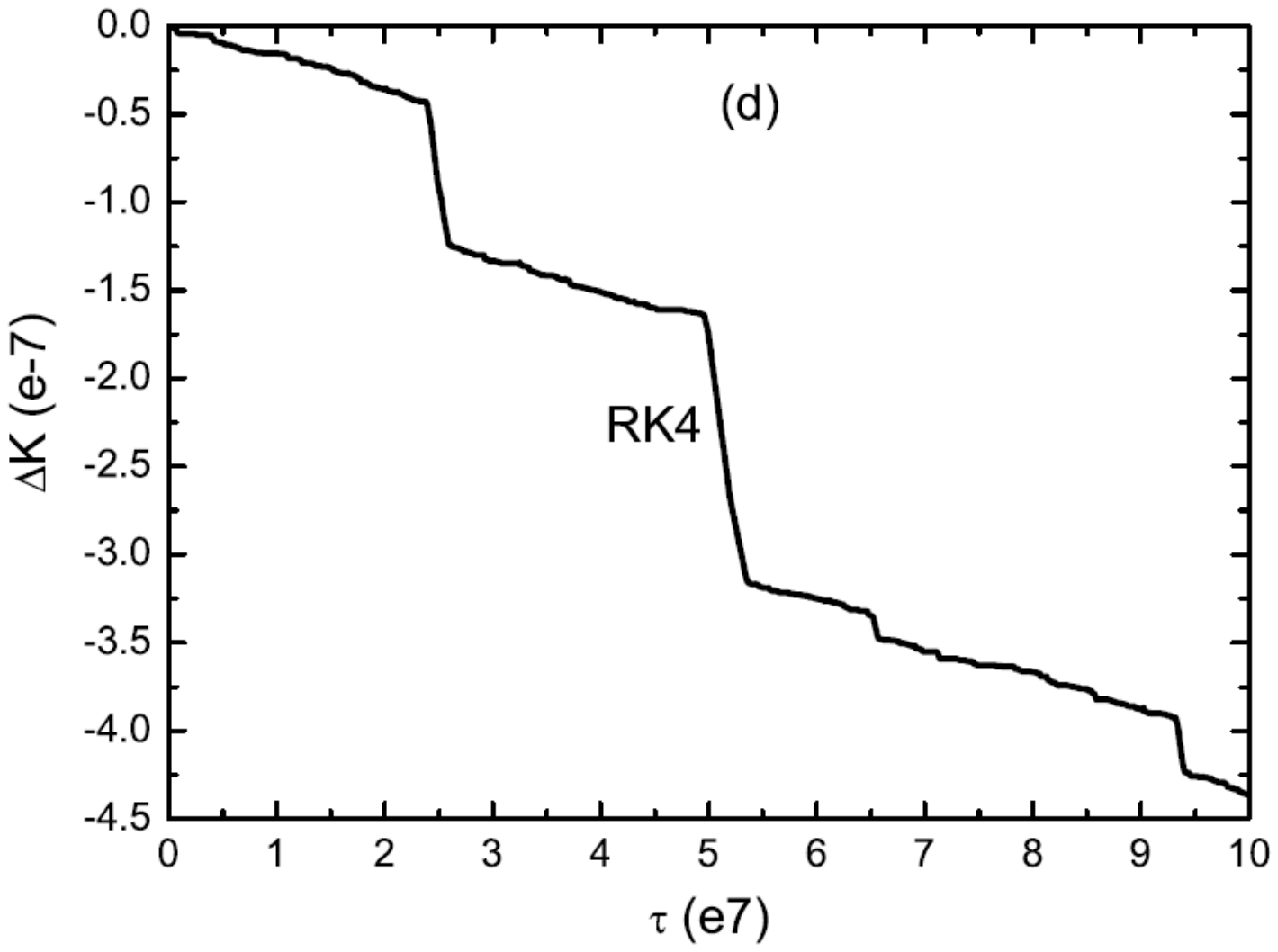}
\caption{Same as Figure 1, but an external magnetic field with
parameter $\beta=8.9\times 10^{-4}$ is included within the
vicinity of the black hole. (a) Poincar\'{e} sections. Orbit 1 is
still a regular torus, Orbit 2 has many islands, and Orbit 3 is
strongly chaotic. (b)-(d) Hamiltonian errors $\Delta K=1+2K$ from
Equation (18) for the algorithms solving the three orbits in the
system (15).}
 \label{Fig2}}
\end{figure*}

\end{document}